\documentclass{ichep04}

\begin{document}

\renewcommand{\d}{{\mathrm d}}
\renewcommand{\bar}[1]{\overline{#1}}

\title{NuTeV Anomaly \& Strange-Antistrange Asymmetry}

\author{Bo-Qiang Ma}

\address{Department of Physics, Peking University, Beijing 100871, China\\E-mail: mabq@phy.pku.edu.cn}



\twocolumn[\maketitle\abstract{ The NuTeV Collaboration reported a
value of $\sin^{2}\theta_{w}$ measured in neutrino-nucleon deep
inelastic scattering, and found that the value is three standard
deviations from the world average value of other electroweak
measurements. If this result cannot be explained within
conventional physics, it must imply some novel physics beyond the
standard model. We report the correction from the asymmetric
strange-antistrange sea by using both the light-cone baryon-meson
fluctuation model and the chiral quark model, and show that a
significant part of the NuTeV anomaly can be explained by the
strange-antistrange asymmetry. }]


The NuTeV Collaboration\cite{zell02} at Fermilab has measured the
value of the Weinberg angle (weak mixing angle)
$\sin^{2}\theta_{w}$ in deep inelastic scattering~(DIS) on nuclear
target with both neutrino and antineutrino beams. Having
considered and examined various source of systematic errors, the
NuTeV Collaboration reported the value:
$$\sin^{2}\theta_{w}=0.2277\pm0.0013~(\mbox{stat})\pm0.0009~(\mbox{syst}),$$
which is three standard deviations from the value
$\sin^{2}\theta_{w}=0.2227\pm0.0004$ measured in other electroweak
processes. As $\theta_{w}$ is one of the important quantities in
the standard model, this observation by NuTeV has received
attention by the physics society. This deviation, or NuTeV anomaly
as people called, could be an indication for new physics beyond
standard model, if it cannot be understood by a reasonable effect
within the standard model.

The NuTeV Collaboration measured the value of $\sin^{2}\theta_{w}$
by using the ratio of neutrino neutral-current and charged-current
cross sections on iron\cite{zell02}. This procedure is closely
related to the Paschos-Wolfenstein~(PW) relation\cite{pash73}:
\begin{equation}
R^{-}=\frac{\sigma^{{\nu}N}_{NC}-\sigma^{\overline{\nu}N}_{NC}}{\sigma^{{\nu}N}_{CC}
-\sigma^{\overline{\nu}N}_{CC}}=\frac{1}{2}-\sin^{2}\theta_{w},
\label{ratio}
\end{equation}
which is based on the assumptions of charge symmetry, isoscalar
target, and strange-antistrange symmetry of the nucleon sea.
There have been a number of corrections considered for the PW
relation, for example: charge symmetry violation\cite{lt03},
neutron excess\cite{k02}, nuclear effect\cite{ksy02},
strange-antistrange asymmetry\cite{cs03,dm04,dxm04}, and also
source for physics beyond standard model\cite{d02}. In this talk,
I will report on the effect due to the strange-antistrange
asymmetry by using both the light-cone baryon-meson fluctuation
model\cite{bm97} and the chiral model
model\cite{Weinberg,Manohar-Georgi}, based on the collaborated
works with Ding\cite{dm04} and also with Ding and Xu\cite{dxm04}.

Among various sources, it is necessary to pay particular attention
to the strange-antistrange asymmetry, which brings the correction
to the PW relation\cite{dm04}
\begin{equation}
R^{-}_{N}=\frac{\sigma^{\nu
N}_{NC}-\sigma^{\bar{\nu}N}_{NC}}{\sigma^{\nu
N}_{CC}-\sigma^{\bar{\nu}N}_{CC}} = R^{-}-\delta
R^{-}_{s},\label{correction}
\end{equation}
where $\delta R^{-}_{s}$ is the correction term
\begin{equation}
\delta
R^{-}_{s}=(1-\frac{7}{3}\sin^{2}\theta_{w})\frac{S^{-}}{Q_v+3
S^{-}},\label{rs}
\end{equation}
where $S^{-}\equiv\int^{1}_{0} x[s(x)-\bar{s}(x)]\textmd{d}x$ and
$Q_v \equiv\int^{1}_{0} x[u_{v}(x)+d_{v}(x)]\textmd{d} x$. A
common assumption about the strange sea is that the $s$ and
$\bar{s}$ distributions are symmetric, but in fact this is
established neither theoretically nor experimentally. It has been
argued recently that there is a strange-antistrange asymmetry in
perturbative QCD at three-loops\cite{pQCD}, although this
perturbative source can only contribute trivially to the NuTeV
anomaly. Possible manifestations of nonperturbative effects for
the strange-antistrange asymmetry have been discussed along with
some phenomenological
explanations\cite{bm97,st87,bw92,hss96,cm98,cs99}. Also there have
been some experimental analyses\cite{b95,sbr97,a97,bpz00}, which
suggest the $s$-$\bar{s}$ asymmetry of the nucleon sea.

It is still controversial whether the strange-antistrange
asymmetry can account for the NuTeV anomaly\cite{o03}. Cao and
Signal\cite{cs03} reexamined the strange-antistrange asymmetry
using the meson cloud model\cite{st87} and concluded that the
second moment $S^{-}\equiv\int^{1}_{0}
x[s(x)-\bar{s}(x)]\textmd{d}x$ is fairly small and unlikely to
affect the NuTeV extraction of $\sin^{2}\theta_{w}$. Oppositely,
Brodsky and I\cite{bm97} proposed a light-cone baryon-meson
fluctuation model to describe the $s(x)-\bar{s}(x)$ distributions
and found a significantly different case from what obtained by
using the meson cloud model\cite{st87,hss96}, as has been
illustrated recently by Ding and I\cite{dm04}. Also, Szczurek {\it
et al.}\cite{sbf96} suggested that the effect of SU(3)$_{f}$
symmetry violation may be specially important in understanding the
strangeness content of the nucleon within the effective chiral
quark model, and compared their results with those of the
traditional meson cloud model qualitatively.

I first present the results by Ding and I\cite{dm04} in the
light-cone baryon-meson fluctuation model. In the light-cone
formalism,
the hadronic wave function can be expressed by a series
of light-cone wave functions multiplied by the Fock states, for
example, the proton wave function can be written as
\begin{eqnarray}
   \left|p\right\rangle=&\left|uud\right\rangle\Psi_{uud/p}+ \left|uudg\right\rangle\Psi_{uudg/p}
\nonumber
\\
   &+\sum_{q\bar{q}}
   \left|uudq \bar{q}\right\rangle\Psi_{uudq\bar{q}/p}+\cdots.
\end{eqnarray}
Brodsky and I made an approximation\cite{bm97}, which suggests
that the intrinsic sea part of the proton function can be
expressed as a sum of meson-baryon Fock states. For example:
$P(uuds\bar{s})=K^{+}(u\bar{s})+\Lambda(uds)$ for the intrinsic
strange sea, the higher Fock states are less important, the $ud$
in $\Lambda$ serves as a spectator in the quark-spectator
model\cite{ma}, for which we choose
\begin{eqnarray}
   \Psi_{D}(x,\mathbf{k}_{\perp})=A_{D}\exp(-M^{2}/8\alpha^{2}_{D}),\label{si}
\end{eqnarray}
\begin{eqnarray}
 \Psi_{D}(x,\mathbf{k}_{\perp})=A_{D}(1+M^{2}/\alpha^{2}_{D})^{-P},\label{psi}
 \end{eqnarray}
where $\Psi_{D}(x,\mathbf{k}_{\perp})$, is a two-body wave
function which is a function of invariant masses for meson-baryon
state:
\begin{eqnarray}
  M^{2}=\frac{m^{2}_{1}+\mathbf{k}^{2}_{\bot}}{x}+\frac{m^{2}_{2}+\mathbf{k}^{2}_{\bot}}{1-x},
\end{eqnarray}
where $\mathbf{k}_{\perp}$ is the initial quark transversal
momentum, $m_{1}$ and $m_{2}$ are the masses for quark $q$ and
spectator $D$, $\alpha_{D}$ sets the characteristic internal
momentum scale, and $P$ is the power constant which is chosen as
$P=3.5$ here. The momentum distribution of the intrinsic $s$ and
$\bar{s}$ in the $K^{+}\Lambda$ state can be modelled from the
two-level convolution formula:
\begin{eqnarray}
   s(x)&=&\int^{1}_{x}\frac{\d y}{y}f_{\Lambda/K^{+}\Lambda}(y)q_{s/\Lambda}(x/y),\nonumber\\
   \bar{s}(x)&=&\int^{1}_{x}\frac{\d y}{y}f_{K^{+}/K^{+}\Lambda}(y)q_{\bar{s}/K^{+}}(x/y),
\end{eqnarray}
where $f_{\Lambda/K^{+}\Lambda}(y)$, $f_{K^{+}/K^{+}\Lambda}(y)$
are the probabilities of finding $\Lambda, K^{+}$ in the
$K^{+}\Lambda$ state with the light-cone momentum fraction $y$,
and $q_{s/\Lambda}(x/y)$, $q_{\bar{s}/K^{+}}(x/y)$ are the
probabilities of finding $s$, $\bar{s}$ quarks in $\Lambda, K^{+}$
state with the light-cone momentum fraction $x/y$.
Two wave function models, the Gaussian type and the power-law
type, are adopted\cite{bm97} to evaluate the asymmetry of
strange-antistrange sea, and almost identical distributions of
$s$-$\bar{s}$ are obtained in the nucleon sea. In this work, we
also consider the two types of wave functions, Eqs.~(\ref{si}) and
(\ref{psi}).

\begin{figure}
\includegraphics[width=6.8cm]{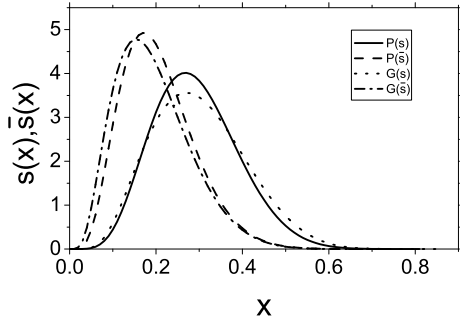}
\caption{\small Distributions for $s(x)$ and $\bar{s}(x)$ in  the
light-cone baryon-meson fluctuation model. $P(s)$ ($G(s)$) is the
$s$ distribution with the power-law wave function (the Gaussian
wave function) and $P(\bar{s}$) ($G(\bar{s}$)) is the
corresponding $\bar s$ distribution.} \label{ssbar}
\end{figure}

The $u$ and $d$ valence quark distributions in the proton are
calculated by using the quark-diquark model\cite{ma}. The
unpolarized valence quark distribution in the proton is
\begin{eqnarray}
   u_{V}(x)&=&\frac{1}{2}a_{S}(x)+\frac{1}{6}a_{V}(x),\nonumber\\
   d_{V}(x)&=&\frac{1}{3}a_{V}(x),
\end{eqnarray}
where $a_{D}(x)$ ($D=S$ or $V$, with $S$ standing for scalar
diquark Fock state and $V$ standing for vector diquark state)
denotes that the amplitude for the quark $q$ is scattered while
the spectator is in diquark state $D$, 
and can be
written as:
\begin{eqnarray}
   a_{D}(x)\propto\int[\d\mathbf{k}_{\bot}]\bigg{|}\Psi_{D}(x,\mathbf{k}_{\bot})\bigg{|}^{2}.
\end{eqnarray}
The values of parameters $\alpha_{D}$, $m_{q}$, and $ m_{D}$ can
be adjusted by fitting the hardonic properties. For light-flavor
quarks, we simple choose $m_{q}=330$~MeV, $\alpha_{D}=330$~MeV,
$m_{S}=600$~MeV, $m_{V}=900$~MeV and
$m_{s}=m_{\bar{s}}=480$~MeV\cite{bm97}. Because the fluctuation
functions were normalized to 1 in Ref.\cite{bm97}, we can obtain
the different distributions for $s$ and $\bar{s}$ in the nucleon.
In the same way, we can get the distributions of the $u$ and $d$
valence quarks, for which the integrated amplitude $\int_0^1 \d x
\,a_D(x)$ must be normalized to 3 in a spectator model
\cite{ma,ma2}. Assuming isospin symmetry, we can get the valence
distributions in the nucleon which implies $N=(p+n)/2$
\begin{eqnarray}
  u^{N}_{V}(x)=\frac{1}{2}\left[\frac{1}{2}a_{S}(x)+\frac{1}{2}a_{V}(x)\right],\nonumber\\
  d^{N}_{V}(x)=\frac{1}{2}\left[\frac{1}{2}a_{S}(x)+\frac{1}{2}a_{V}(x)\right].
\end{eqnarray}
Thus, using this model, we obtain the distributions of $s$ and
$\bar{s}$ in the nucleon sea. The numerical result is given in
Fig.~\ref{ssbar}. One can find that $s<\bar{s}$ as $x<0.235$,
$s>\bar{s}$ as $x>0.235$, this result is opposite to the
prediction from the meson cloud model\cite{cs03}. 
From
Eq.~(\ref{correction}), one can find that a shift of $\delta
R^{-}_{s}$ should lead to a shift in the $R^{-}$, which affect the
extraction of $\sin^{2}\theta_{w}$, Eq.~(\ref{rs}). The result of
our calculation is 0.0042$<S^{-}<$0.0106 (0.0035$<S^{-}<$0.0087)
for the Gaussian wave function (for the power-law wave function),
which corresponds to $P_{K^{+}\Lambda}$=4\%, 10\%. Hence,
$0.0017<\delta R^{-}_{s}<0.0041$ ($0.0014<\delta
R^{-}_{s}<0.0034$), for the Gaussian wave function (the power-law
wave function). The shift in $\sin^{2}\theta_{w}$ can reduce the
discrepancy from 0.005 to 0.0033 (0.0036) ($P_{K^{+}\Lambda}$=4\%)
or 0.0009 (0.0016) ($P_{K^{+}\Lambda}$=10\%).

In the above, our attention is on the distributions of $s(x)$ and
$\bar{s}(x)$, and on calculating the second moment $S^{-}$ by
using the light-cone baryon-meson fluctuation model\cite{bm97}. We
find that the $s$-$\bar{s}$ asymmetry can remove the NuTeV anomaly
by about 30--80\%.

\begin{figure}
\scalebox{0.62}{\includegraphics[0,16][307,212]{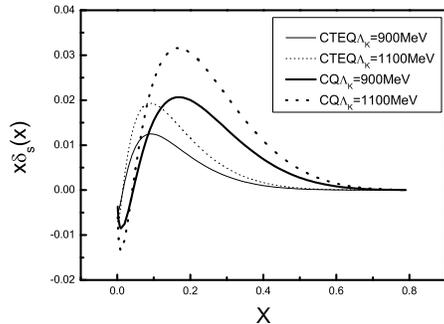}}
\caption{\small The distributions of
$x\left[s(x)-\bar{s}(x)\right]$ in the chiral quark model, with
inputs of valence quark distributions from both constituent quark
(CQ) model~(thick curves) and CTEQ6 parametrization~(thin curves),
and the cut-off parameter $\Lambda_{K}=900$~MeV (solid curves) and
1100~MeV (dashed curves ).}
\end{figure}

 A~~further study by Ding, Xu and I\cite{dxm04} by using chiral quark model also shows that this
strange-antistrange asymmetry has a significant contribution to
the PW relation and can explain the anomaly without sensitivity to
input parameters. The chiral symmetry at high energy scale and it
breaking at low energy scale are the basic properties of QCD. The
chiral quark model, established by Weinberg\cite{Weinberg}, and
developed by Manohar and Georgi\cite{Manohar-Georgi}, has been
widely accepted by the hadron physics society as an effective
theory of QCD at low energy scale. This model has also a number of
phenomenological applications, such as to explain the light-flavor
sea asymmetry of $u$ and $d$ sea quarks\cite{ehq92}, and also to
understand the proton spin problem\cite{cl95}. In this new work,
we provide a new success to understand the NuTeV anomaly with the
chiral quark model without sensitivity on parameters. We find that
the effect due to strange-antistrange asymmetry can bring a
significant contribution to the NuTeV anomaly of about 60--100\%
with reasonable parameters without sensitivity to different inputs
of constituent quark distributions. This may imply that the NuTeV
anomaly can be considered as a phenomenological support to the
strange-antistrange asymmetry of the nucleon sea. Thus it is
important to make a precision measurement of the distributions of
$s(x)$ and $\bar{s}(x)$ in the nucleon more carefully in future
experiments.

\end{document}